\documentclass[runningheads]{llncs}
\usepackage[T1]{fontenc}
\usepackage[utf8]{inputenc}
\usepackage{graphicx}
\usepackage{booktabs}
\usepackage{amsmath}
\usepackage{amssymb}
\usepackage{fancyvrb}
\usepackage{multirow}
\usepackage{tabularx}
\usepackage{marvosym}
\DefineVerbatimEnvironment{Code}{Verbatim}{fontsize=\scriptsize,frame=single,framesep=4pt}
\begin{document}
\title{Harness TTS: Towards Context-Aware Expressive Speech Synthesis with Harness Layer}
\titlerunning{Harness TTS}

\author{Shengfan Shen\inst{1,2} \and Di Wu\inst{2,3} \and Xingchen Song\inst{2,3} \and Dinghao Zhou\inst{2,3} \and
Pengyu Cheng\inst{2} \and Sixiang Lyu\inst{2} \and Jian Luan\inst{2} \and Shuai Wang\inst{1,3}\textsuperscript{\Letter}}
\authorrunning{S. Shen et al.}
\institute{Nanjing University, China \\ \email{shenshengfan@hnu.edu.cn,shuaiwang@nju.edu.cn} \and MiLM Plus, Xiaomi Inc., China \and WeNet Open Source Community}
\maketitle
\begin{abstract}
Expressive speech synthesis for voice assistants requires flexible style control that adapts to explicit requests
and broader interaction context. We propose Harness TTS, a lightweight control layer that wraps around a TTS engine to externalize and govern its expressive behavior. It reformulates style control as closed-set prompt-tool routing: offline, a compact registry of stylistic prompt tools is constructed with structured metadata; online, an LLM planner selects the appropriate tool based on a priority-aware observation schema, and the TTS executor synthesizes speech using the corresponding prompt audio. We evaluate Harness TTS on both routing and synthesis tasks. In routing, Qwen3-4B achieves Top-1 accuracies of 74.3\%, 43.0\%, and 64.6\% on explicit, implicit, and conflict subsets. For synthesis, experiments on CosyVoice3 and VoxCPM2 show that Harness TTS outperforms instruction-only control, achieving higher instruction-following win rates (margins of 23.1–35.6 points on CosyVoice3 and 13.8–20.0 points on VoxCPM2) and improving UTMOSv2 scores by 0.11–0.38. Moreover, the 4B planner delivers its first tool recommendation in under 50 ms in standard mode, introducing negligible latency for real-time interaction. These results demonstrate that equipping TTS engines with a dedicated Harness layer offers a practical, auditable, and context-aware solution for voice assistant expression control.

\end{abstract}
\section{Introduction}

Despite the rapid progress of large-scale generative TTS models in producing highly natural and emotionally diverse speech~\cite{du2025cosyvoice,zhou2026indextts2,yang2024instructtts}, effectively harnessing these expressive capabilities in practical voice assistants remains challenging. Zero-shot TTS typically relies on a fixed reference prompt for speaker imitation, which limits flexible style adaptation. Instruct TTS introduces natural language descriptions for style control but often yields unstable outputs under ambiguous requests. More importantly, existing paradigms primarily focus on the immediate input, lacking a systematic mechanism to incorporate broader contextual signals—such as user profiles, interaction scenarios, and environmental conditions—that are essential for personalized voice assistants.

We argue that the major bottleneck is not the generative capacity of modern TTS models, but the absence of a systematic interface layer that bridges contextual understanding and speech generation. Drawing inspiration from harness engineering, where external orchestration layers organize information, constrain behaviors, and improve model usability~\cite{zhou2026externalization}, we introduce a lightweight Harness layer for TTS. Rather than modifying the underlying TTS engine, our approach augments existing models by externalizing expressive decision-making: the Harness interprets contextual signals, selects an appropriate expressive prompt, and delegates speech generation to the TTS backend. This separation of concerns enables more stable, interpretable, and context-aware expressive synthesis without altering the TTS backend. In practical voice assistant scenarios, maintaining a curated collection of high-quality expressive prompts for a target speaker—and selecting the most suitable one according to each interaction context—has emerged as a pragmatic and reliable strategy. Under this setting, the key challenge shifts from generating arbitrary styles to reliably matching contextual requirements with available expressive resources. However, existing solutions—heuristic rules, or embedding-based retrieval—often fail to handle implicit preferences, resolve conflicting requirements, or make priority-aware decisions.

To address this challenge, we instantiate the Harness layer as a lightweight decision-making component on top of the TTS engine. Offline, curated expressive prompts are organized into a structured tool registry with descriptive metadata. During online inference, an LLM-based planner receives a multi-source observation, resolves potential conflicts through a predefined priority policy, and selects an appropriate prompt. The selected prompt is then executed by the TTS engine to synthesize speech. By interposing this Harness layer between contextual understanding and speech generation, we transform expressive requests into closed-set prompt-tool routing, allowing existing TTS engines to more reliably realize their inherent expressive potential.

Our core contributions are as follows:
\begin{itemize}
\item \textbf{A lightweight Harness layer for expressive TTS}: We introduce an intermediate layer that augments existing TTS engines with expressive prompt selection, enabling more stable and interpretable style realization without modifying the underlying generator.
\item \textbf{A context-grounded planning mechanism}: We design a multi-level observation schema and a priority-based conflict resolution strategy, allowing the Harness layer to integrate diverse contextual signals and make reliable expressive decisions.
\item \textbf{Comprehensive evaluation}: Through routing and synthesis benchmarks, we demonstrate that Harness-guided synthesis improves instruction following and naturalness over instruction-only control while maintaining competitive speaker consistency.
\end{itemize}

\section{Related Work}
\label{sec:related}

\paragraph{Zero-shot TTS}
Zero-shot TTS has evolved from timbre cloning to fine-grained style control. Early work such as VALL-E~\cite{10842513} demonstrated the feasibility of cloning voices from short prompt audio via in-context learning. Subsequent models including NaturalSpeech 2~\cite{shen2024naturalspeech}, Voicebox~\cite{le2023voicebox}, and CosyVoice 2~\cite{du2024cosyvoice} further improved naturalness and speaker similarity through large-scale training. More recent systems like MegaTTS 2~\cite{jiang2024mega} and IndexTTS 2~\cite{zhou2026indextts2} explore the disentanglement of speech attributes including timbre, prosody, and emotion, enabling independent control over these dimensions. Despite these advances, zero-shot TTS typically relies on a fixed reference prompt at runtime. This design limits its flexibility for real-time adaptation, motivating the need for a selection-based mechanism that can dynamically choose among multiple candidate prompts.

\paragraph{Instruct TTS}
Instruct TTS enable flexible style control through natural language descriptions. PromptTTS~\cite{guo2023prompttts} first validated the feasibility of controlling acoustic attributes using text prompts. Later work like InstructTTS~\cite{yang2024instructtts} adopted longer descriptive sentences for richer style specification. OV-InstructTTS~\cite{ren2026ov} introduced a two-step reasoning framework that infers acoustic attributes from open-ended instructions, further improving instruction following. Additionally, CosyVoice 3~\cite{du2025cosyvoice} and VoxCPM2~\cite{zhou2026voxcpm2} allow users to adjust speaking style via natural language while preserving consistent timbre through audio cloning. These methods offer the advantage of open-vocabulary expressiveness. However, mapping free-form instructions to acoustic realization can still be challenging due to ambiguous descriptions and the stochastic nature of generative inference, making it difficult to guarantee stable behavior in production deployments.

\paragraph{Context-Aware Conversational TTS}
Speech synthesis in dialogue scenarios requires models to go beyond the current utterance and understand the emotional and prosodic dynamics conveyed in the conversation history. FCTalker~\cite{hu2024fctalker} captures contextual dependencies at both the word and sentence levels. M²-CTTS~\cite{xue2023m} further incorporates both textual and acoustic modalities to model prosodic variations more comprehensively. CapTalk~\cite{su2026captalk} introduces speaker-level descriptions and Chain-of-Thought reasoning to plan prosodic attributes turn by turn. The ISCSLP 2026 CoT-TTS Challenge~\cite{xue2026iscslp} tasks participants with reasoning about speaking styles from dialogue histories, further promoting this direction. Nevertheless, current conversational TTS primarily rely on short-term dialogue history, lacking broader signals such as user profiles, scenario types, or ambient cues, which limits their adaptability in complex real-world scenarios.

\section{Method}

\begin{figure}[htbp]
    \centering
    \includegraphics[width=\linewidth]{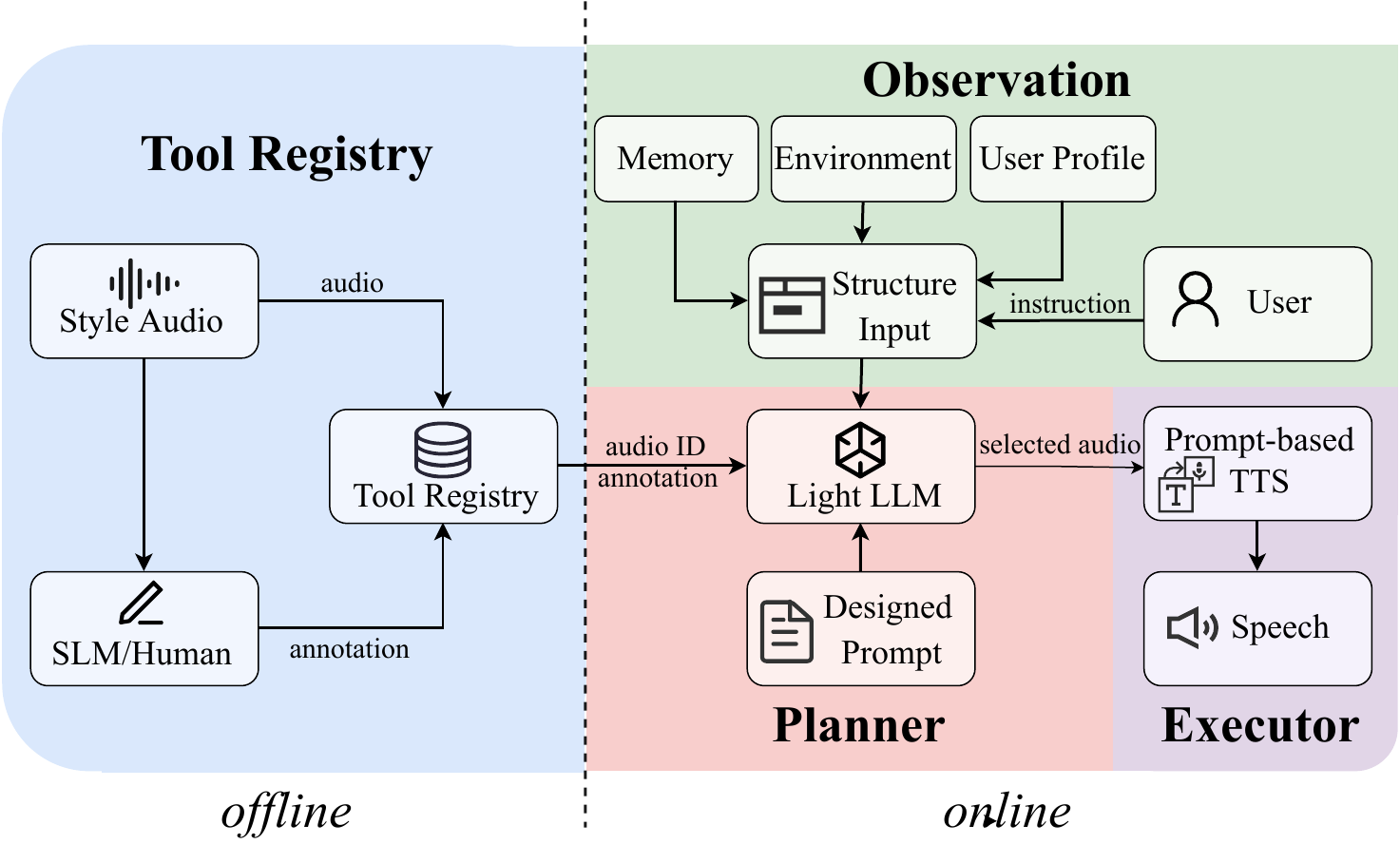}
    \caption{Overview of the Harness TTS architecture. The offline stage constructs the tool registry. The online stage consists of three steps: observation, planning, and execution. The planner selects a tool from the registry given the structured observation, and the executor synthesizes speech using the corresponding prompt.}
    \label{fig:overview}
\end{figure}

\subsection{Problem Formulation}

We approach context-aware expressive TTS from the perspective of harness engineering, which advocates externalizing control logic and decision-making from the core generative model into an orchestration layer. As illustrated in Figure~\ref{fig:overview}, our Harness layer wraps around a TTS engine $\mathcal{E}$ and operates through a closed-set tool routing framework. Instead of navigating an unbounded, continuous acoustic parameter space, we bound the decision space to a finite, pre-registered tool registry $\mathcal{T}$. Each tool in $\mathcal{T}$ corresponds to a specific audio prompt and is uniquely identified by an \texttt{audio\_id}. 

This closed-set design provides two critical advantages: (1) \textbf{constrained decision space}, which strictly limits the output scope to enhance controllability and system stability; and (2) \textbf{traceable routing}, which ensures every stylistic decision is fully interpretable and debuggable. During inference, given the structured contextual observation $o$ and tool metadata, a compact LLM planner $\pi$ outputs a ranked list of candidate tools. The executor $\mathcal{E}$ then conditions its speech generation on the prompt audio associated with the top-ranked tool ID.

\subsection{Tool Registry}

\begin{figure}[htbp]
    \centering
    \includegraphics[width=\linewidth]{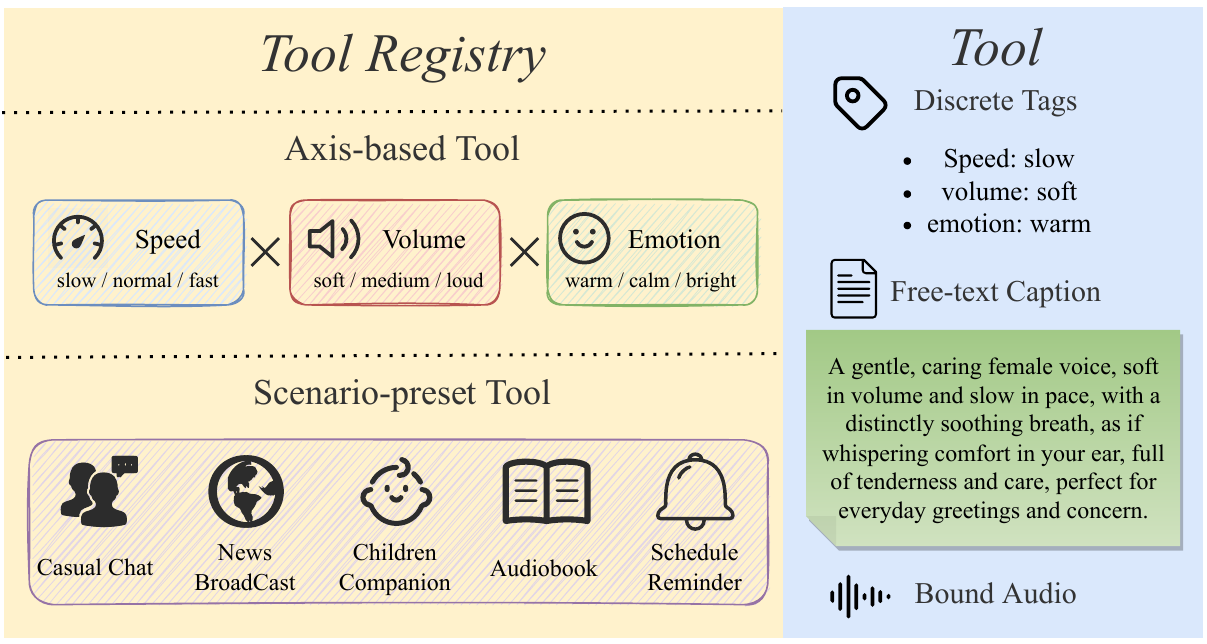}
    \caption{Categorical construction of the tool registry (left) and content of each tool entry (right). The registry consists of two categories: Axis-based tools and Scenario-preset tools. Each tool contains discrete tags, a free-text caption, and the bound audio.}
    \label{fig:tool_registry}
\end{figure}

The tool registry is the foundational component of the Harness layer, defining the set of available expressive controls that govern the TTS engine's output. Unlike approaches that build large-scale multi-speaker voice libraries, we construct a compact, style-oriented prompt library for a single target speaker. This design narrows the planner's decision space while meeting the practical requirement for speaker consistency in voice assistants. Each tool in the registry $\mathcal{T}$ is associated with a prompt audio and structured textual metadata, allowing the planner to select suitable tools without accessing raw audio. Figure~\ref{fig:tool_registry} illustrates the registry structure: the left panel presents the two tool categories, and the right panel details the contents of each entry.

\subsubsection{Tool Categories: Axis-Based and Scenario-Preset Tools}

The registry comprises two complementary categories: \textbf{axis-based tools} and \textbf{scenario-preset tools}. Axis-based tools enable systematic control over low-level acoustic attributes across three primary axes---speed, volume, and emotion. Combinations of these attributes yield diverse expressive variants, supporting fine-grained adjustments to the TTS output. Conversely, for frequently encountered interaction scenarios, scenario-preset tools provide a dedicated style audio for each case, offering directly applicable expressive patterns for common use cases.

\subsubsection{Dual-Annotation Structure}

Each tool comprises a bound audio and a structured description that enables the Harness planner to reason about its function. To balance precision and flexibility, the description follows a dual-annotation format combining discrete tags and free-text captions. Specifically, \textbf{discrete tags} cover low-level acoustic attributes such as speed, volume, and emotion to facilitate precise property matching during routing. Complementing these, a \textbf{free-text caption} provides a natural-language summary of the tool's applicable scenarios and auditory style, delivering broader semantic and contextual associations beyond the discrete tags.

\subsection{Observation Schema and Priority Policy}
\label{sec:obs}

Existing expressive TTS systems primarily control speaking style based on the current utterance and explicit user instructions. In practice, however, voice assistant scenarios involve additional contextual signals—such as user profiles, interaction scenarios, and ambient cues—that substantially influence style expression. To capture this richer context, our observation module aggregates heterogeneous upstream information and structures it into five decision layers: system defaults, user profile, scenario, implicit intent cues, and explicit instructions. This schema enables the Harness layer to reason beyond \textit{what is being said} to \textit{under what scenario, for whom, and with what intended expression}.

\begin{table}[htbp]
\centering
\scriptsize
\renewcommand{\arraystretch}{1.2}
\begin{tabular}{@{}c m{0.20\linewidth} m{0.60\linewidth}@{}}
\toprule
Priority & Field & Description \\
\midrule
5 & Explicit instruction & Direct style request (e.g., ``speak slowly and gently''). \\
4 & Implicit intent & Contextual cues (time, ambient noise, dialogue history, user emotion) that implicitly reveal user needs. \\
3 & Scene & Dialogue scenario (e.g., navigation, bedtime story, news broadcast, reminder, or safety alert). \\
2 & User profile & User information such as age group, preferred and avoided speaking styles. \\
1 & System default & Default persona style used when no stronger evidence is present. \\
\bottomrule
\end{tabular}
\caption{Observation hierarchy within the Harness layer. Higher-priority fields override lower-priority fields in case of conflicts.}
\label{tab:obs-hierarchy}
\end{table}

These five fields follow a predefined priority order, as detailed in Table~\ref{tab:obs-hierarchy}. This priority policy serves as a key embedded decision rule, ensuring that the planner resolves conflicting signals in a predictable and interpretable manner. The design adheres to a practical interaction principle: explicit and immediate user requirements override implicit, scenario-based, or default preferences. Lower-priority fields supply background context, while higher-priority fields dominate when they convey more specific or immediate instructions.

\subsection{LLM-based Planner}
\label{sec:planner}

The planner is the decision-making core of the Harness layer, implemented as a text-only router built on a lightweight LLM. This design externalizes style selection from the TTS engine into the Harness, leveraging LLM reasoning for tool selection while maintaining low inference latency. Its input comprises three components: (i)~the tool registry $\mathcal{T}$, (ii)~the structured observation $o$, and (iii)~a routing prompt encoding the priority policy. To improve the planner's understanding of routing criteria and output format adherence, we incorporate few-shot examples covering three representative scenarios: explicit instruction, implicit intent, and conflicting contextual requirements. We further experiment with a Chain-of-Thought (CoT)~\cite{wei2022chain} variant, where the planner produces structured output with \texttt{<reason>} and \texttt{<audio\_id>} tags for each candidate, enforcing explicit reasoning before each selection at the cost of additional inference latency.

\subsection{Executor: Prompt-Based TTS}
\label{sec:executor}

The executor is the governed component of the Harness system---the underlying TTS engine that the Harness layer controls. Given the target utterance text \(x\) and the selected \texttt{audio\_id}, it retrieves the corresponding prompt audio from the registry and synthesizes the response. The executor exposes a simple interface: the planner determines only the prompt identity, while the executor maps the selected tool to the corresponding acoustic conditioning signal. This modular design cleanly separates contextual decision-making, handled by the Harness, from speech generation, performed by the engine, enabling seamless integration of various prompt-based TTS backends. Ideally, the TTS executor is more than a generic zero-shot TTS engine: it should be fine-tuned on the target speaker and capable of cloning that speaker's voice across diverse styles. This ensures both stable style rendering and consistent timbre preservation.

\section{Experiments}

We evaluate the proposed Harness layer along three research questions:

\textbf{(RQ1)} Whether the Harness planner selects appropriate expressive tools from the registry under structured contextual observations;

\textbf{(RQ2)} Whether the Harness layer, built on a lightweight LLM, meets the latency requirements for real-time interactive scenarios;

\textbf{(RQ3)} Whether Harness-guided selection improves expressive realization in downstream speech synthesis over instruction-only control.

To address these questions, we design two complementary experiments. The routing task targets RQ1 and RQ2 by assessing the planner's decision-making capability and inference latency. The synthesis task targets RQ3 by evaluating the complete Harness pipeline in end-to-end speech generation.

\subsection{Routing Task}

\paragraph{Data Construction}
We construct an evaluation dataset to assess the planner's decision-making capability under various routing scenarios. Since the planner relies solely on textual metadata without accessing audio, we create a simulated registry of 42 tools annotated with captions and tags only. This provides a reasonably diverse decision space for routing evaluation. The registry comprises 27 axis-based tools covering all \(3 \times 3 \times 3\) combinations of speed, volume, and emotion, and 15 scenario-preset tools for common voice-assistant use cases. Based on this registry, we construct three subsets, each targeting a distinct aspect of routing behavior:

\begin{itemize}
    \item \textbf{Explicit Subset.} Cases where user instructions directly specify the desired speech style attributes. This subset tests the planner's ability to follow explicit style requirements.
    \item \textbf{Implicit Subset.} Cases where the preferred tool must be inferred from contextual cues---such as ambient noise, time, user emotion, and interaction scenario---rather than from explicit directives. User input, if present, conveys only user states or environmental feedback. This subset evaluates the planner's capacity for contextual reasoning.
    \item \textbf{Conflict Subset.} Cases where contextual fields intentionally conflict with the explicit user instruction. Correct routing follows the predefined priority policy and prioritizes explicit user requirements over other contextual signals. This subset examines whether the planner adheres to the specified conflict resolution rule.
\end{itemize}

For each subset, we prompt Gemini-2.5-Pro~\cite{comanici2025gemini} to generate 5 test cases per tool, following manually designed templates that encourage the corresponding routing patterns. This yields 210 samples per subset.

\paragraph{Systems Compared}
We employ Gemini-2.5-Pro as the teacher model to produce ranked tool labels for every test case. Student planners include Qwen3-0.6B, Qwen3-1.7B, and Qwen3-4B~\cite{yang2025qwen3}. We compare our LLM-based planners against two types of non-planner baselines:

\begin{itemize}
\item \textbf{Keyword baseline}: We design manually specified keywords for scenario-preset tools and for each attribute value of axis-based tools. During inference, keyword matching is performed over the observation text. Matched keywords determine the corresponding scenario or attribute values, while unmatched attributes are randomly filled to simulate a rule-based system without semantic reasoning.
\item \textbf{Retrieval baselines}: We use Qwen3-Embedding~\cite{zhang2025qwen3} to encode tool captions and observation text, and retrieve the top-five tools using cosine similarity. We experiment with two model sizes, denoted as Retrieval-0.6B and Retrieval-4B, corresponding to Qwen3-Embedding-0.6B and Qwen3-Embedding-4B, respectively. This baseline performs semantic matching without explicitly modeling field priorities or conflict resolution.
\end{itemize}

\paragraph{Accuracy Metrics}
We report teacher-alignment metrics, prefixed with ``T-'' to indicate that the reference is the teacher model rather than human labels. Let \(g_1\) and \(G_5\) denote the teacher's top-ranked tool and top-five set, respectively, while \(P_K\) represents the planner's top-\(K\) predictions. We evaluate performance using four metrics:
\textbf{T-Exact@1} ($\mathbf{1}[P_1 = g_1]$) checks for an exact match with the teacher's top choice; 
\textbf{T-Contain@K} ($\mathbf{1}[g_1 \in P_K]$) measures if the teacher's top tool appears within the first $K$ predictions; 
\textbf{T-Recall@5} ($|P_5 \cap G_5| / |G_5|$) calculates the recall against the teacher's top-5 set; and 
\textbf{T-Jaccard@5} ($|P_5 \cap G_5| / |P_5 \cup G_5|$) quantifies the Jaccard similarity between the two top-5 sets.

\paragraph{Latency Metrics}
We measure planner-only inference latency on an RTX 5090 GPU with vLLM~\cite{kwon2023efficient} serving. \textbf{P50/P95 Full} denote the time (ms) to produce the complete ranked list of five candidates. \textbf{P50/P95 First-ID} denote the time until the first valid \texttt{audio\_id} is generated. We report First-ID latency separately because synthesis can begin as soon as the top tool is known, without waiting for the full list.

\subsection{Synthesis Task}

We evaluate synthesis quality by comparing two control conditions on both CosyVoice3~\cite{du2025cosyvoice} and VoxCPM2~\cite{zhou2026voxcpm2}: 
(1) \textbf{Harness}, which utilizes Qwen3-4B as planner; and 
(2) \textbf{Instruct}, where the TTS engine directly receives the instruction while conditioned on a default reference audio of the target speaker.

\paragraph{Data Construction}
The voice library used in this experiment contains 25 audio clips in total: 14 axis-based tools, 10 scenario-preset tools, and 1 default tool. We select 27 style instruction tags covering two main categories: acoustic control and application scenarios. For each tag, we generate 5 utterances with different textual content using an LLM, resulting in 135 test cases. Since the Instruct condition only accepts user instructions and cannot utilize other contextual information, we construct the test cases solely around user instructions. For the Harness condition, we set all other fields in the structured observation to default neutral values to ensure comparability.

\paragraph{Executor Metrics}
We use three automatic metrics to evaluate synthesis quality: 
\textbf{Instruction Following Win Rate (Inst. win\%)} uses Gemini-3.1-Pro to assess how well the synthesized speech follows the style instruction. The judge is provided with the instruction and two randomly shuffled audio samples (Harness vs. Instruct) to eliminate position bias, and is prompted to first generate a reason and then output a preference.
\textbf{UTMOSv2}~\cite{baba2024t05} provides a no-reference prediction of speech naturalness. 
\textbf{Speaker Stability (Spk-Stab)} measures timbre consistency under slight instruction variations by calculating the average pairwise speaker similarity among all synthesized utterances within each instruction tag using WeSpeaker~\cite{wang2023wespeaker}.

\section{Results}

\subsection{RQ1: Routing Accuracy}

\begin{table}[ht]
\centering
\caption{Performance on the explicit subset. Bold indicates the best performance, and underline indicates the second-best performance across all methods.}
\label{tab:explicit}
\scriptsize
\setlength{\tabcolsep}{4pt}
\begin{tabular}{@{}lccccc@{}}
\toprule
Method & T-Exact@1 & T-Contain@3 & T-Contain@5 & T-Recall@5 & T-Jaccard@5 \\
\midrule
Keyword          & 0.376 & 0.533 & 0.576 & 0.392 & 0.274 \\
Retrieval-0.6B   & 0.329 & 0.529 & 0.643 & 0.331 & 0.229 \\
Retrieval-4B     & 0.348 & 0.543 & 0.662 & 0.223 & 0.215 \\
Qwen3-0.6B w/ CoT  & 0.214 & 0.362 & 0.390 & 0.275 & 0.202 \\
Qwen3-0.6B w/o CoT & 0.033 & 0.090 & 0.133 & 0.207 & 0.131 \\
Qwen3-1.7B w/ CoT  & 0.543 & 0.619 & 0.643 & 0.374 & 0.270 \\
Qwen3-1.7B w/o CoT & 0.348 & 0.510 & 0.610 & 0.420 & 0.296 \\
Qwen3-4B w/ CoT    & \textbf{0.743} & \textbf{0.886} & \textbf{0.914} & \underline{0.540} & \textbf{0.431} \\
Qwen3-4B w/o CoT   & \underline{0.614} & \underline{0.767} & \underline{0.833} & \textbf{0.559} & \underline{0.415} \\
\bottomrule
\end{tabular}
\end{table}

\paragraph{Explicit Subset.}
When users provide explicit style instructions, Qwen3-4B with CoT achieves the highest T-Exact@1 (74.3\%) in Table~\ref{tab:explicit}, showing a large teacher-alignment gain over the best retrieval baseline (Retrieval-4B: 34.8\%). Model scale significantly impacts decision accuracy: the 4B variant outperforms the 1.7B model (54.3\%) and the 0.6B model (21.4\%) by large margins. We attribute this gap to the smaller models' difficulty in following instructions and reasoning under lengthy contexts, which occasionally leads to hallucinated or repeated tool outputs, further degrading effective accuracy. CoT prompting improves performance across all model sizes, albeit at the cost of increased inference latency. Retrieval baselines can locate the correct tool within their top-5 candidates (T-Contain@5: up to 66.2\%) but exhibit weaker ranking precision, as reflected by their lower T-Exact@1 scores. This gap underscores the value of incorporating an LLM-based decision core within the Harness layer, rather than relying solely on embedding similarity for selection.

\begin{table}[ht]
\centering
\caption{Performance on the implicit subset.}
\label{tab:implicit}
\scriptsize
\setlength{\tabcolsep}{4pt}
\begin{tabular}{@{}lccccc@{}}
\toprule
Method & T-Exact@1 & T-Contain@3 & T-Contain@5 & T-Recall@5 & T-Jaccard@5 \\
\midrule
Keyword          & 0.207 & 0.267 & 0.385 & 0.439 & 0.306 \\
Retrieval-0.6B   & 0.193 & 0.296 & 0.385 & 0.316 & 0.206 \\
Retrieval-4B     & 0.259 & 0.393 & 0.459 & 0.255 & 0.135 \\
Qwen3-0.6B w/ CoT  & 0.259 & 0.378 & 0.393 & 0.265 & 0.194 \\
Qwen3-0.6B w/o CoT & 0.081 & 0.237 & 0.259 & 0.176 & 0.121 \\
Qwen3-1.7B w/ CoT  & 0.252 & 0.444 & 0.504 & 0.373 & 0.276 \\
Qwen3-1.7B w/o CoT & \underline{0.333} & 0.467 & 0.585 & 0.409 & 0.281 \\
Qwen3-4B w/ CoT    & \textbf{0.430} & \textbf{0.652} & \textbf{0.756} & \underline{0.505} & \textbf{0.394} \\
Qwen3-4B w/o CoT   & 0.311 & \underline{0.496} & \underline{0.607} & \textbf{0.507} & \underline{0.372} \\
\bottomrule
\end{tabular}
\end{table}

\paragraph{Implicit Subset.}
When tool selection depends on contextual cues, all methods exhibit lower T-Exact@1 (Table~\ref{tab:implicit}) compared with the explicit subset. This overall accuracy drop stems partly from task ambiguity: implicit cues are less constraining than explicit instructions, naturally lowering the ceiling for agreement with the teacher's top-1 prediction. Despite this, Qwen3-4B with CoT achieves the highest T-Exact@1 (43.0\%), outperforming Retrieval-4B (25.9\%) by a considerable margin. In such cases, retrieval baselines, which rely on surface-level semantic similarity, struggle with cross-semantic inferences---e.g., mapping ``late night'' to a ``soft and gentle'' style. The LLM planner, in contrast, demonstrates a clear advantage in candidate coverage, with its top prediction falling within the teacher's top-5 set 75.6\% of the time. This suggests that LLM-based planning is more adaptable in scenarios where direct semantic matching proves insufficient.

\begin{table}[ht]
\centering
\caption{Performance on the conflict subset.}
\label{tab:conflict}
\scriptsize
\setlength{\tabcolsep}{4pt}
\begin{tabular}{@{}lccccc@{}}
\toprule
Method & T-Exact@1 & T-Contain@3 & T-Contain@5 & T-Recall@5 & T-Jaccard@5 \\
\midrule
Keyword          & 0.286 & 0.364 & 0.398 & 0.236 & 0.157 \\
Retrieval-0.6B   & 0.233 & 0.364 & 0.456 & 0.245 & 0.157 \\
Retrieval-4B     & 0.199 & 0.364 & 0.456 & 0.223 & 0.137 \\
Qwen3-0.6B w/ CoT  & 0.165 & 0.262 & 0.311 & 0.200 & 0.144 \\
Qwen3-0.6B w/o CoT & 0.034 & 0.097 & 0.136 & 0.137 & 0.087 \\
Qwen3-1.7B w/ CoT  & 0.490 & 0.558 & 0.568 & 0.283 & 0.195 \\
Qwen3-1.7B w/o CoT & 0.194 & 0.325 & 0.388 & 0.250 & 0.164 \\
Qwen3-4B w/ CoT    & \textbf{0.646} & \textbf{0.840} & \textbf{0.883} & \textbf{0.450} & \textbf{0.335} \\
Qwen3-4B w/o CoT   & \underline{0.495} & \underline{0.631} & \underline{0.709} & \underline{0.395} & \underline{0.273} \\
\bottomrule
\end{tabular}
\end{table}

\paragraph{Conflict Subset.}
When contextual fields are deliberately set to contradict the user's explicit instruction, correct routing requires the planner to enforce the priority hierarchy. As shown in Table~\ref{tab:conflict}, Qwen3-4B with CoT achieves 64.6\% T-Exact@1, substantially outperforming Retrieval-4B (19.9\%)---the largest gap among all three subsets. This performance disparity stems from a fundamental design difference: as an embedding-based matcher, the flat embedding retrieval baseline cannot resolve conflicts between signals, whereas our LLM planner is explicitly designed to identify such conflicts and invoke the priority policy.

\subsection{RQ2: Planner Latency}

\begin{table}[ht]
\centering
\caption{Planner inference latency (ms) across different methods. P50 and P95 denote median and 95th-percentile latencies.}
\label{tab:latency}
\scriptsize
\setlength{\tabcolsep}{4pt}
\begin{tabular}{@{}lccccc@{}}
\toprule
Method & P50 Full & P95 Full & P50 First-ID & P95 First-ID \\
\midrule
Retrieval-0.6B   & 26.5 & 27.1 & 26.5 & 27.1 \\
Retrieval-4B     & 54.6 & 56.1 & 54.6 & 56.1 \\
Qwen3-0.6B w/o CoT & 57.5 & 66.9 & 10.6 & 12.6 \\
Qwen3-1.7B w/o CoT & 92.2 & 102.0 & 17.5 & 31.8 \\
Qwen3-4B w/o CoT   & 179.9 & 206.0 & 34.4 & 41.2 \\
Qwen3-0.6B w/ CoT  & 258.9 & 401.5 & 65.4 & 80.4 \\
Qwen3-1.7B w/ CoT  & 459.2 & 537.4 & 104.8 & 135.1 \\
Qwen3-4B w/ CoT    & 979.8 & 1191.7 & 216.3 & 323.6 \\
\bottomrule
\end{tabular}
\end{table}

Retrieval baselines achieve the lowest full-list latency (P95 Full: 27.1--56.1 ms), as they only perform query embedding and similarity computation against pre-computed tool embeddings. For LLM planners without CoT, P95 Full latency ranges from 66.9 ms (0.6B) to 206.0 ms (4B). More importantly, P95 First-ID latency is substantially lower: the 4B model generates its first valid \texttt{audio\_id} within 41.2 ms, which is comparable to retrieval baselines. Since most deployments require only the top-ranked tool to initiate synthesis, First-ID latency is the more practically relevant metric for real-time interaction. Under this metric, the 4B model without CoT introduces only marginal latency, well within the requirements of interactive applications. CoT prompting increases latency, yet its First-ID latency remains acceptable for applications without stringent real-time constraints.

\subsection{RQ3: Synthesis Quality}

\begin{figure}[htbp]
    \centering
    \includegraphics[width=\linewidth]{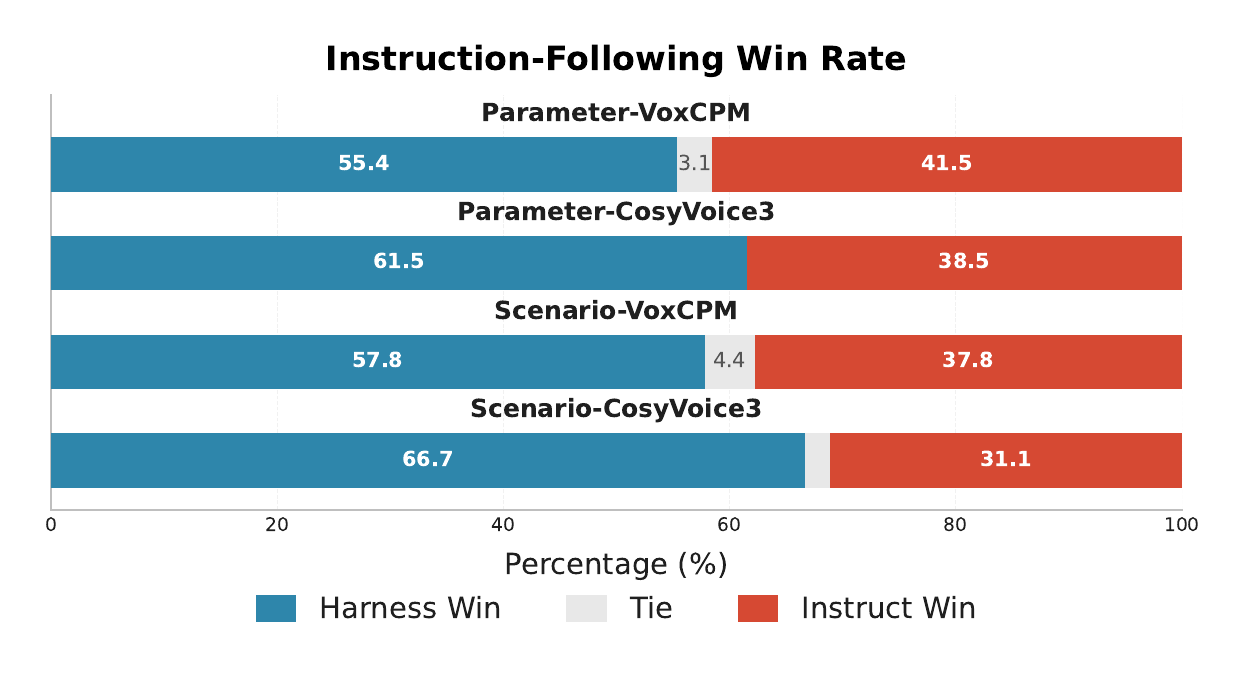}
    \caption{Instruction-following win rates under the Harness and Instruct conditions, with CosyVoice3 and VoxCPM2 as the TTS executor. Results are shown separately for parameter and scenario categories.}
    \label{fig:executor_winrate}
\end{figure}

\paragraph{Instruction-Following Win Rate.}
As shown in Figure~\ref{fig:executor_winrate}, Harness consistently achieves higher instruction-following win rates than Instruct across all four conditions. On VoxCPM2, Harness outperforms Instruct by 13.8 and 20.0 percentage points on the parameter and scenario subsets, respectively. The improvement is more pronounced on CosyVoice3, with margins of 23.1 percentage points on the parameter subset and 35.6 percentage points on the scenario subset. For both backends, the gains on the scenario subset are larger than those on the parameter subset, suggesting that instruction-only control struggles to realize high-level scenario descriptions. In contrast, Harness can either directly match the described scenario to a corresponding preset tool or, when no direct match exists, infer the underlying acoustic attributes and compose an appropriate axis-based candidate. This flexibility makes Harness particularly advantageous for holistic, high-level style instructions.

\begin{figure}[htbp]
    \centering
    \includegraphics[width=\linewidth]{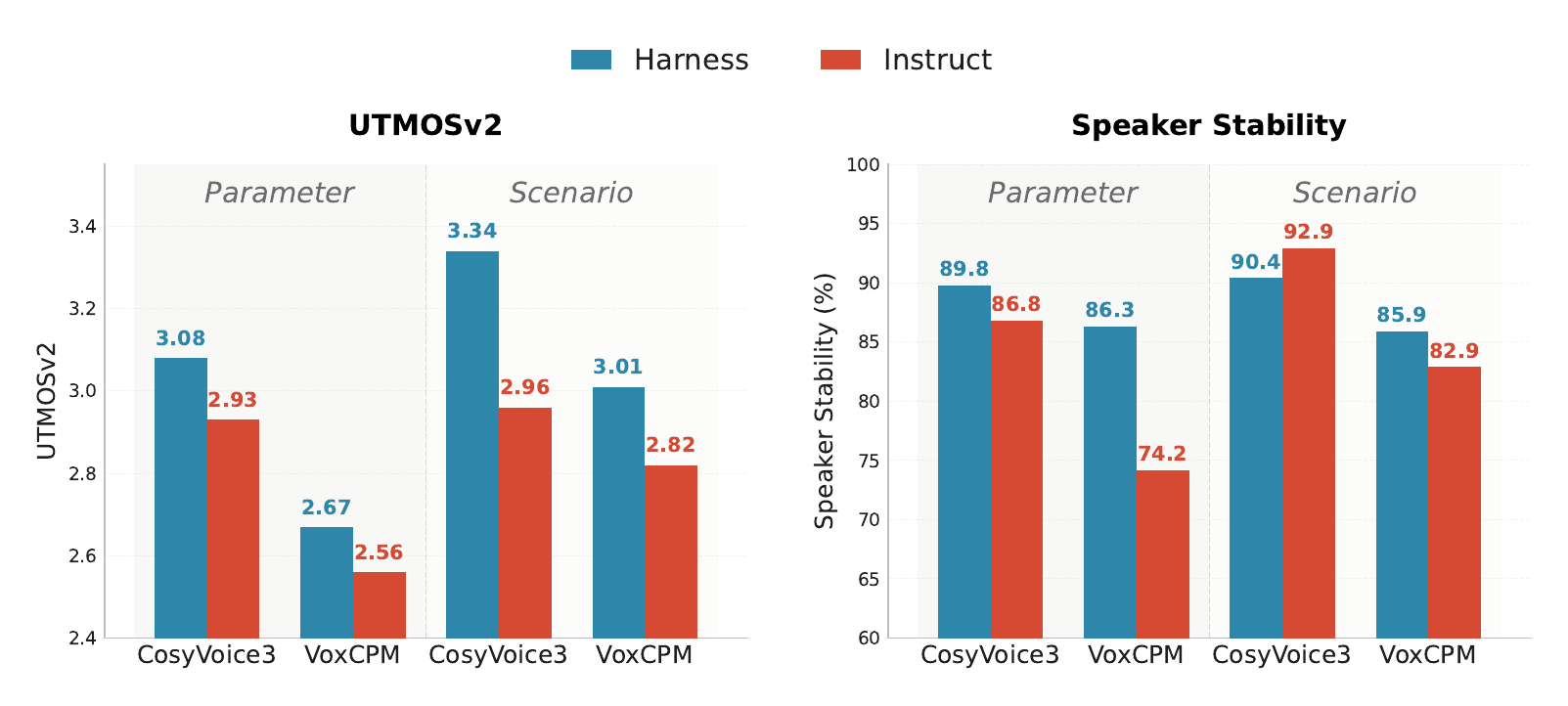}
    \caption{UTMOSv2 and Speaker Stability under Harness and Instruct conditions, with CosyVoice3 and VoxCPM2 as the TTS executor.}
    \label{fig:executor_utmos_spk}
\end{figure}

\paragraph{Naturalness and Speaker Stability.}
Figure~\ref{fig:executor_utmos_spk} presents the UTMOSv2 and speaker stability results. For naturalness, Harness outperforms Instruct across all four conditions, indicating improved perceptual quality. For speaker stability, VoxCPM2 shows consistent gains with Harness on both subsets, suggesting more stable speaker characteristics under Harness. On CosyVoice3, Harness achieves slightly higher stability on the parameter subset (89.8 vs.~86.8), while Instruct shows a marginally higher score on the scenario subset (92.9 vs.~90.4). This latter result, however, should be interpreted alongside the instruction-following outcomes. As shown in Figure~\ref{fig:executor_winrate}, CosyVoice3-Instruct yields substantially lower Inst. win\% on the scenario subset, indicating that its higher speaker stability may be partly attributable to poorer instruction adherence, which limits stylistic variation and artificially inflates timbre consistency. In contrast, Harness successfully delivers the required style while maintaining competitive speaker stability. Overall, these results suggest that Harness enhances both instruction following and naturalness without sacrificing speaker consistency.

\section{Limitations}

Our study has several limitations. First, while CoT prompting improves routing accuracy, it incurs additional inference latency that may affect real-time interaction; distilling the teacher model into a smaller student model could offer a more favorable accuracy--speed trade-off. Second, our evaluation relies on LLM-based teachers and judges rather than human annotations; thus, the reported metrics should be interpreted as automated benchmarks that await subjective validation. Third, all test data are generated by LLMs rather than collected from real-world user logs, so the generalization of the proposed Harness layer to naturally occurring conversational scenarios remains to be investigated.

\section{Conclusion}

We presented Harness TTS, a lightweight Harness layer that wraps around a TTS engine to enable context-grounded, auditable, and stable expressive control for voice assistants. Drawing inspiration from harness engineering, our approach externalizes the decision-making logic---interpreting multi-source contextual signals, resolving conflicting requirements, and selecting appropriate styles---from the TTS engine into a control layer. This design preserves the engine's generative fidelity while enhancing it with observability, interpretability, and priority-aware decision-making. Experimental results demonstrate strong routing accuracy across explicit, implicit, and conflict scenarios, outperforming retrieval-based baselines. Synthesis evaluations further confirm that Harness-guided synthesis improves instruction following and naturalness over instruction-only control, while maintaining competitive speaker consistency. These findings validate the effectiveness of augmenting TTS engines with a dedicated Harness layer, offering a practical and extensible pathway toward personalized, context-aware voice assistants.

\bibliographystyle{splncs04}   
\bibliography{reference}    

@ARTICLE{10842513,
  author={Chen, Sanyuan and Wang, Chengyi and Wu, Yu and Zhang, Ziqiang and Zhou, Long and Liu, Shujie and Chen, Zhuo and Liu, Yanqing and Wang, Huaming and Li, Jinyu and He, Lei and Zhao, Sheng and Wei, Furu},
  journal={IEEE Transactions on Audio, Speech and Language Processing}, 
  title={Neural Codec Language Models are Zero-Shot Text to Speech Synthesizers}, 
  year={2025},
  volume={33},
  number={},
  pages={705-718},
  doi={10.1109/TASLPRO.2025.3530270}
}

@inproceedings{shen2024naturalspeech,
  title={Naturalspeech 2: Latent diffusion models are natural and zero-shot speech and singing synthesizers},
  author={Shen, Kai and Ju, Zeqian and Tan, Xu and Liu, Eric and Leng, Yichong and He, Lei and Qin, Tao and Bian, Jiang and others},
  booktitle={International conference on learning representations},
  volume={2024},
  pages={698--722},
  year={2024}
}

@article{le2023voicebox,
  title={Voicebox: Text-guided multilingual universal speech generation at scale},
  author={Le, Matthew and Vyas, Apoorv and Shi, Bowen and Karrer, Brian and Sari, Leda and Moritz, Rashel and Williamson, Mary and Manohar, Vimal and Adi, Yossi and Mahadeokar, Jay and others},
  journal={Advances in neural information processing systems},
  volume={36},
  pages={14005--14034},
  year={2023}
}

@article{du2024cosyvoice,
  title={Cosyvoice 2: Scalable streaming speech synthesis with large language models},
  author={Du, Zhihao and Wang, Yuxuan and Chen, Qian and Shi, Xian and Lv, Xiang and Zhao, Tianyu and Gao, Zhifu and Yang, Yexin and Gao, Changfeng and Wang, Hui and others},
  journal={arXiv preprint arXiv:2412.10117},
  year={2024}
}

@inproceedings{jiang2024mega,
  title={Mega-tts 2: Boosting prompting mechanisms for zero-shot speech synthesis},
  author={Jiang, Ziyue and Liu, Jinglin and Ren, Yi and He, Jinzheng and Ye, Zhenhui and Ji, Shengpeng and Yang, Qian and Zhang, Chen and Wei, Pengfei and Wang, Chunfeng and others},
  booktitle={International Conference on Learning Representations},
  volume={2024},
  pages={57919--57939},
  year={2024}
}

@inproceedings{zhou2026indextts2,
  title={Indextts2: A breakthrough in emotionally expressive and duration-controlled auto-regressive zero-shot text-to-speech},
  author={Zhou, Siyi and Zhou, Yiquan and He, Yi and Zhou, Xun and Wang, Jinchao and Deng, Wei and Shu, Jingchen},
  booktitle={Proceedings of the AAAI Conference on Artificial Intelligence},
  volume={40},
  number={41},
  pages={35139--35148},
  year={2026}
}

@inproceedings{guo2023prompttts,
  title={Prompttts: Controllable text-to-speech with text descriptions},
  author={Guo, Zhifang and Leng, Yichong and Wu, Yihan and Zhao, Sheng and Tan, Xu},
  booktitle={ICASSP 2023-2023 IEEE International Conference on Acoustics, Speech and Signal Processing (ICASSP)},
  pages={1--5},
  year={2023},
  organization={IEEE}
}

@article{yang2024instructtts,
  title={Instructtts: Modelling expressive tts in discrete latent space with natural language style prompt},
  author={Yang, Dongchao and Liu, Songxiang and Huang, Rongjie and Weng, Chao and Meng, Helen},
  journal={IEEE/ACM Transactions on Audio, Speech, and Language Processing},
  volume={32},
  pages={2913--2925},
  year={2024},
  publisher={IEEE}
}

@article{ren2026ov,
  title={OV-InstructTTS: Towards Open-Vocabulary Instruct Text-to-Speech},
  author={Ren, Yong and Yi, Jiangyan and Tao, Jianhua and Sun, Haiyang and Wen, Zhengqi and Gu, Hao and Xu, Le and Bai, Ye},
  journal={arXiv preprint arXiv:2601.01459},
  year={2026}
}

@article{du2025cosyvoice,
  title={Cosyvoice 3: Towards in-the-wild speech generation via scaling-up and post-training},
  author={Du, Zhihao and Gao, Changfeng and Wang, Yuxuan and Yu, Fan and Zhao, Tianyu and Wang, Hao and Lv, Xiang and Wang, Hui and Ni, Chongjia and Shi, Xian and others},
  journal={arXiv preprint arXiv:2505.17589},
  year={2025}
}

@article{zhou2026voxcpm2,
  title={VoxCPM2 Technical Report},
  author={Zhou, Yixuan and Zeng, Guoyang and Liu, Xin and Li, Xiang and Yu, Renjie and Gui, Jiancheng and Wu, Jiaheng and Wang, Ziyang and Shen, Xudong and Ye, Runchuan and others},
  journal={arXiv preprint arXiv:2606.06928},
  year={2026}
}

@inproceedings{hu2024fctalker,
  title={Fctalker: Fine and coarse grained context modeling for expressive conversational speech synthesis},
  author={Hu, Yifan and Liu, Rui and Gao, Guanglai and Li, Haizhou},
  booktitle={2024 IEEE 14th International Symposium on Chinese Spoken Language Processing (ISCSLP)},
  pages={299--303},
  year={2024},
  organization={IEEE}
}

@inproceedings{xue2023m,
  title={M 2-ctts: End-to-end multi-scale multi-modal conversational text-to-speech synthesis},
  author={Xue, Jinlong and Deng, Yayue and Wang, Fengping and Li, Ya and Gao, Yingming and Tao, Jianhua and Sun, Jianqing and Liang, Jiaen},
  booktitle={ICASSP 2023-2023 IEEE International Conference on Acoustics, Speech and Signal Processing (ICASSP)},
  pages={1--5},
  year={2023},
  organization={IEEE}
}

@article{su2026captalk,
  title={Captalk: Unified voice design for single-utterance and dialogue speech generation},
  author={Su, Xiaosu and Sun, Zihan and Jia, Peilei and Gao, Jun},
  journal={arXiv preprint arXiv:2604.08363},
  year={2026}
}

@article{xue2026iscslp,
  title={ISCSLP 2026 CoT-TTS Challenge: Chain-of-Thought Reasoning for Context-Aware Text-to-Speech},
  author={Xue, Wei and Feng, Junlan and Zhang, Shilei and Wang, Yue and Yang, Ruosong and Liu, Bei and Xue, Liumeng and Cheng, Sitong and Pan, Jiahao and Bian, Weizhen and others},
  journal={arXiv preprint arXiv:2606.21933},
  year={2026}
}

@article{yang2025qwen3,
  title={Qwen3 technical report},
  author={Yang, An and Li, Anfeng and Yang, Baosong and Zhang, Beichen and Hui, Binyuan and Zheng, Bo and Yu, Bowen and Gao, Chang and Huang, Chengen and Lv, Chenxu and others},
  journal={arXiv preprint arXiv:2505.09388},
  year={2025}
}

@inproceedings{kwon2023efficient,
  title={Efficient memory management for large language model serving with pagedattention},
  author={Kwon, Woosuk and Li, Zhuohan and Zhuang, Siyuan and Sheng, Ying and Zheng, Lianmin and Yu, Cody Hao and Gonzalez, Joseph and Zhang, Hao and Stoica, Ion},
  booktitle={Proceedings of the 29th symposium on operating systems principles},
  pages={611--626},
  year={2023}
}

@article{zhang2025qwen3,
  title={Qwen3 embedding: Advancing text embedding and reranking through foundation models},
  author={Zhang, Yanzhao and Li, Mingxin and Long, Dingkun and Zhang, Xin and Lin, Huan and Yang, Baosong and Xie, Pengjun and Yang, An and Liu, Dayiheng and Lin, Junyang and others},
  journal={arXiv preprint arXiv:2506.05176},
  year={2025}
}

@article{comanici2025gemini,
  title={Gemini 2.5: Pushing the frontier with advanced reasoning, multimodality, long context, and next generation agentic capabilities},
  author={Comanici, Gheorghe and Bieber, Eric and Schaekermann, Mike and Pasupat, Ice and Sachdeva, Noveen and Dhillon, Inderjit and Blistein, Marcel and Ram, Ori and Zhang, Dan and Rosen, Evan and others},
  journal={arXiv preprint arXiv:2507.06261},
  year={2025}
}

@article{wei2022chain,
  title={Chain-of-thought prompting elicits reasoning in large language models},
  author={Wei, Jason and Wang, Xuezhi and Schuurmans, Dale and Bosma, Maarten and Xia, Fei and Chi, Ed and Le, Quoc V and Zhou, Denny and others},
  journal={Advances in neural information processing systems},
  volume={35},
  pages={24824--24837},
  year={2022}
}

@inproceedings{wang2023wespeaker,
  title={Wespeaker: A research and production oriented speaker embedding learning toolkit},
  author={Wang, Hongji and Liang, Chengdong and Wang, Shuai and Chen, Zhengyang and Zhang, Binbin and Xiang, Xu and Deng, Yanlei and Qian, Yanmin},
  booktitle={ICASSP 2023-2023 IEEE International Conference on Acoustics, Speech and Signal Processing (ICASSP)},
  pages={1--5},
  year={2023},
  organization={IEEE}
}

@inproceedings{baba2024t05,
  title={The t05 system for the voicemos challenge 2024: Transfer learning from deep image classifier to naturalness mos prediction of high-quality synthetic speech},
  author={Baba, Kaito and Nakata, Wataru and Saito, Yuki and Saruwatari, Hiroshi},
  booktitle={2024 IEEE Spoken Language Technology Workshop (SLT)},
  pages={818--824},
  year={2024},
  organization={IEEE}
}

@article{zhou2026externalization,
  title={Externalization in llm agents: A unified review of memory, skills, protocols and harness engineering},
  author={Zhou, Chenyu and Chai, Huacan and Chen, Wenteng and Guo, Zihan and Shan, Rong and Song, Yuanyi and Xu, Tianyi and Yang, Yingxuan and Yu, Aofan and Zhang, Weiming and others},
  journal={arXiv preprint arXiv:2604.08224},
  year={2026}
}

\end{document}